%% file: main.tex
\pdfoutput=1

\documentclass[11pt]{article}

\usepackage[preprint]{acl}

\usepackage[utf8]{inputenc} 
\usepackage{times}    
\usepackage{hyperref}       
\usepackage{url}            
\usepackage{booktabs}       
\usepackage{amsfonts}       
\usepackage{nicefrac}       
\usepackage{microtype}      
\usepackage{lipsum}
\usepackage{subcaption}
\usepackage{tabularx}
\usepackage{xcolor}
\usepackage{multirow}
\usepackage{amsmath}
\usepackage{fancyhdr}       
\usepackage{graphicx}       
\graphicspath{{media/}}     

\title{MiniELM: A Lightweight and Adaptive Query Rewriting Framework for E-Commerce Search Optimization}

\author{Duy A. Nguyen$^{1*\dagger}$, Rishi Kesav Mohan$^{1*}$, Van Yang$^{1*}$, \\\textbf{Pritom Saha Akash}$^{1*}$, \textbf{Kevin Chen-Chuan Chang}$^{1*}$ \\
$^{1}$Siebel School of Computing and Data Science \\
$^{*}$University of Illinois, Urbana-Champaign, US \\
$^\dagger$VinUni-Illinois Smart Health Center, VinUniversity, Vietnam \\
\texttt{\{duyan2,rkmohan2,sy65,pakash2,kcchang\}@illinois.edu}
}

\begin{document}
\pagestyle{empty}
\maketitle

\begin{abstract}
\input{sections/abstract}
\end{abstract}

\input{sections/introduction}
\input{sections/literature}

\input{sections/problem_definition}

\input{sections/method}
\input{sections/experiment}
\input{sections/discussion_conclusion}

\newpage
\section*{Limitations} 
\label{sec:limitation}
While the current implementation demonstrates significant contributions, there are limitations that require further investigation. MiniELM is currently tailored for English queries, limiting its usability in multilingual e-commerce platforms. Expanding the framework to accommodate multiple languages would improve its generalization. Moreover, while simulated feedback effectively accelerates online adaptation, incorporating real human feedback—or a hybrid approach combining both simulated and real feedback—could further enhance its performance.

\section*{Ethical Considerations and Broader Impact}
MiniELM introduces improvements in query rewriting for e-commerce, but its deployment should be taken with care to avoid potential ethical concerns related to bias and transparency. Since the model learns from historical data, it may reinforce existing biases, favoring popular brands or frequently searched products while underrepresenting niche sellers. Transparency is another key concern, as users and merchants have limited visibility into how and why their queries are rewritten. Without interpretability mechanisms, MiniELM’s query modifications could lead to unintended shifts in search results, affecting user trust and seller visibility. 

Despite these concerns, MiniELM has the potential for significant positive impact on e-commerce search experiences if it is correctly deployed. By bridging lexical gaps and enhancing query diversity, it improves product discoverability, allowing users to find relevant items more easily, even with ambiguous or misspelled queries. This benefits both consumers and smaller sellers, as it enables lesser-known products to surface in search results. Additionally, MiniELM’s adaptive reinforcement learning mechanism ensures that query rewrites evolve with changing trends, reducing reliance on static query expansion rules. For e-commerce platforms, this leads to better search efficiency, increased user engagement, and a more scalable approach to query understanding without costly human annotations.

\section*{Acknowledgment}
This material is based upon work supported by the National Science Foundation IIS 16-19302 and IIS 16-33755, Zhejiang University ZJU Research 083650, IBM-Illinois Center for Cognitive Computing Systems Research (C3SR) and IBM-Illinois Discovery Accelerator Institute (IIDAI), grants from eBay and Microsoft Azure, UIUC OVCR CCIL Planning Grant 434S34, UIUC CSBS Small Grant 434C8U, and UIUC New Frontiers Initiative. Any opinions, findings, conclusions, or recommendations expressed in this publication are those of the author(s) and do not necessarily reflect the views of the funding agencies.

The work of Duy A. Nguyen was supported in part by a PhD fellowship from the VinUni-Illinois Smart Health Center, VinUniversity, Hanoi, Vietnam.

\bibliography{references}

\newpage
\input{sections/appendix}
\end{document}

%% file: sections/abstract.tex
Query rewriting (QR) is a critical technique in e-commerce search, addressing the lexical gap between user queries and product descriptions to enhance search performance. Existing QR approaches typically fall into two categories: discriminative models and generative methods leveraging large language models (LLMs). Discriminative models often struggle with natural language understanding and offer limited flexibility in rewriting, while generative LLMs, despite producing high-quality rewrites, face high inference latency and cost in online settings. These limitations force offline deployment, making them vulnerable to issues like information staleness and semantic drift.
To overcome these challenges, we propose a novel hybrid pipeline for QR that balances efficiency and effectiveness. Our approach combines \textbf{offline knowledge distillation} to create a lightweight but efficient student model with \textbf{online reinforcement learning (RL)} to refine query rewriting dynamically using real-time feedback. A key innovation is the use of LLMs as \textbf{simulated human feedback}, enabling scalable reward signals and cost-effective evaluation without manual annotations. Experimental results on Amazon ESCI dataset demonstrate significant improvements in query relevance, diversity, and adaptability, as well as positive feedback from the LLM simulation.
This work contributes to advancing LLM capabilities for domain-specific applications, offering a robust solution for dynamic and complex e-commerce search environments.

%% file: sections/introduction.tex
\section{Introduction}
\label{sec:intro}
\begin{figure}[t]
    \centering
    \includegraphics[width=\columnwidth]{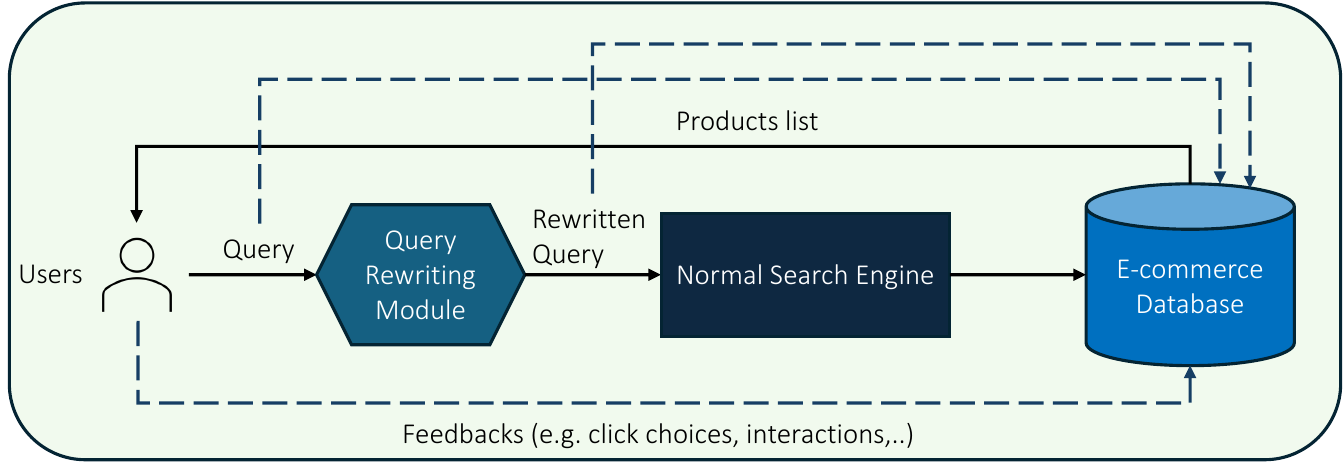}
    \caption{Overview of an E-commerce search pipeline with Query Rewriting module installed.}
    \label{fig:qr}
\end{figure}

\textbf{Context.} Product search is a central component of e-commerce platforms like Amazon or eBay, enabling users to discover relevant items from vast catalogs. In these platforms, users often face challenges when formulating queries, leading to suboptimal search experiences. These challenges are magnified in scenarios where users may not use precise or correct terminology, employ synonyms, or mix languages in their search phrases due to ineptitude of language proficiency. Additionally, the search terms might be misspelled or overly general, making it difficult for traditional search systems to retrieve relevant products. For example, a user may search for ``dress'', which is too broad, while others might input ``summer dress'', ``boho maxi dress'', or ``red evening gown'', each reflecting different intents but lacking clarity without additional context. As e-commerce platforms continue to grow in both scale and diversity, ensuring accurate and relevant product retrieval becomes increasingly difficult, necessitating the need for advanced query rewriting techniques.
query rewriting (QR) refers to the process of transforming an input query into one or more alternative queries that are semantically similar but may be phrased differently, thereby improving the likelihood of retrieving more relevant products. \emph{In the context of e-commerce platforms, effective QR is crucial for bridging the gap between user intent and the diverse ways products can be described in the catalog} (Figure \ref{fig:qr}). 

\textbf{Previous literature.} Query rewriting (QR) methods can be broadly categorized into discriminative and generative approaches. Further details about existing work are provided in Appendix \ref{sec:literature}.

\textbf{Discriminative methods} \cite{xu2017query, mandal2019query, li2022query, shekarpour2017rquery, diaz2016pseudo} focus on reformulating queries by identifying similar terms from a predefined query rewriting set, leveraging sparse retrieval techniques to find relevant products. For example, using traditional Information Retrieval (IR) techniques, a query like ``laptop under 500'' might be rewritten as ``budget laptop'' or ``cheap laptops'' by detecting semantically similar phrases. While computationally efficient, these methods face critical limitations. They often struggle with long-tail queries, where reformulation sets lack appropriate alternatives, leading to inadequate or irrelevant rewrites. Furthermore, their reliance on static, predefined mappings limits flexibility, particularly for queries with complex or ambiguous user intent. Addressing these challenges requires a more dynamic and adaptable approach capable of handling diverse user inputs.

In response to these limitations, \textbf{Generative methods} \cite{agrawal2023rl, qiu2021query, jagerman2023llm}, such as those using Large Language Models (LLMs), have gained popularity due to their superior language understanding and contextual flexibility. By training on extensive corpora of query-reformulation pairs, generative models can produce diverse, contextually relevant rewrites. For instance, an LLM might reformulate the query ``best wireless headphones'' into alternatives like ``top-rated wireless earphones'' or ``best Bluetooth headphones'', potentially enhancing the coverage and relevance of search results. These methods represent a significant leap forward, offering the ability to dynamically generate novel query reformulations without relying on predefined sets.

However, generative methods also have their drawbacks, particularly in real-world e-commerce applications. The large-scale nature of LLMs results in high inference latency and computational costs, making real-time deployment impractical. To mitigate this, LLMs are often deployed in an ``offline'' manner, precomputing query rewrites for popular searches and storing them in cache memory \cite{agrawal2023rl}. While this reduces latency, it introduces issues like information staleness, as the models are not continuously updated to reflect new products, trends, or user behavior. This is especially problematic in e-commerce, where product catalogs and user preferences evolve rapidly, leading to outdated or irrelevant rewrites. These challenges highlight the need for a solution that combines the language ability of LLMs with a compact, efficient, and real-time adaptable framework.

The online deployment of an efficient and effective query rewriting module in e-commerce search systems remains a significant challenge for existing approaches. Ideally, such a module should retain the strong language capabilities of an LLM while being compact, resource-efficient, and practical for real-time deployment.

\textbf{Contribution.} 
In this paper, we propose a novel adaptive query rewriting pipeline that effectively balances efficiency and performance, addressing the limitations of current approaches.

Our solution employs a dual-phase training framework for a large language model (LLM), integrating offline and online training. In the offline phase, we leverage knowledge distillation to create a compact and efficient student model, termed the \textbf{Mini E-commerce Language Model (MiniELM)}, distilled from a large foundation teacher model while preserving semantic fidelity. In the online phase, MiniELM is fine-tuned using reinforcement learning with dynamic reward signals derived from simulated user feedback. This approach not only reduces inference costs but also ensures that the model aligns with and captures relevance, diversity, and user preferences in product retrievals.

A key innovation of our method is the use of \emph{simulated human feedback via LLMs}, replacing resource-intensive manual annotations. This mechanism effectively mimics real-world deployment scenarios while enabling scalable evaluation and continuous model refinement. Additionally, we introduce reward models that assess query rewrites on relevance, diversity, and coverage of user intent, ensuring comprehensive performance metrics. Experimental results on the Amazon ESCI dataset \cite{esci} validate MiniELM’s effectiveness across both offline and online stages, demonstrating its superiority over baseline methods.
In summary, our contributions are as follows:

\begin{itemize} 
\itemsep0em
\item Propose MiniELM, a lightweight and efficient query rewriting model derived through knowledge distillation.
\item Introduce a two-phase training framework integrating offline knowledge distillation and online reinforcement learning.
\item Develop scalable reward models and leverage LLM-based simulated feedback to refine query rewriting dynamically.
\item Validate MiniELM through extensive experiments on the Amazon ESCI dataset, showcasing its effectiveness and superiority.
\end{itemize}

%% file: sections/literature.tex
\section{Related Works}
\label{sec:literature}
\subsection{Discriminative Method}
Discriminative methods frame query rewriting as a retrieval task, expanding original queries with relevant terms using pseudo-relevance feedback, thesaurus-based techniques, and search log-based methods. These approaches represent a progression toward addressing semantic drift, adaptability, and personalization challenges.

Pseudo-relevance feedback methods, such as those by Xu and Croft \cite{xu2017query}, identify expansion terms from top-ranked documents of an initial query, blending global corpus analysis with local feedback. While effective against word mismatches, they are prone to semantic drift from noisy or irrelevant top results, necessitating more stable resources.

Thesaurus-based methods mitigate this instability by using predefined lexical resources like WordNet. Mandal et al. \cite{mandal2019query} advanced this approach with synonym extraction and Boolean query generation, improving recall. However, thesaurus dependency limits adaptability to dynamic trends or rare queries, prompting the need for real-time, user-driven solutions.

Search log-based techniques address these limitations by leveraging user interactions, such as query transitions and clicks, to generate rewrite candidates dynamically. Li et al. \cite{li2022query} demonstrated their adaptability to evolving trends and contextual personalization. Yet, biases toward frequently searched queries hinder their performance on long-tail terms, emphasizing the need for approaches that combine real-time insights with robust language understanding.

These advancements highlight the evolution of discriminative methods toward adaptive and user-informed query rewriting, while still grappling with semantic reliability, trend adaptability, and query diversity.

\subsection{Generative Method}

Generative methods have revolutionized query rewriting by leveraging advanced neural architectures and training paradigms. Prominent approaches include reinforcement learning (RL)-enhanced methods, transformer-based models, and Large Language Model (LLM)-driven techniques. RL-based methods optimize generative models for task-specific goals, such as balancing relevance and diversity, using custom reward functions. Agrawal et al. \cite{agrawal2023rl} demonstrate their ability to align queries with human preferences and maximize product coverage, though scalability and performance on long-tail queries remain challenging.

Transformer-based models, like the cyclic translation framework by Qiu et al. \cite{qiu2021query}, utilize pre-trained architectures to maintain semantic consistency between rewritten and original queries. This approach excels in handling frequent and dynamic queries but depends heavily on the quality of pre-trained models and translation mechanisms.

LLMs, as demonstrated in \citet{jagerman2023llm, llm_1, llm_2, llm_3}, generate semantically rich, diverse query expansions through strategies like zero-shot, few-shot, and Chain-of-Thought prompting. PRF-enhanced prompts further improve contextual understanding, but these models face challenges in fine-tuning for specific goals and impose high resource demands. Product-agent systems, such as those by Zhou et al. \cite{zhou2024agent}, extend LLM capabilities by integrating APIs and knowledge graphs, enabling dynamic query adaptation and addressing standalone LLM limitations.

Generative methods, particularly LLMs, face challenges in real-time e-commerce applications due to high inference latency and computational costs, making them unsuitable for direct online deployment. As a workaround, LLMs are often used in an "offline" manner, where rewritten queries for popular searches are precomputed and cached \cite{agrawal2023rl, jagerman2023llm}. While this reduces latency, it introduces issues of staleness, as offline models are not continuously updated to reflect new products, trends, or user behaviors. In dynamic e-commerce environments, this can result in reformulations that fail to align with evolving trends or updated product categories, ultimately degrading the relevance and quality of search results.

%% file: sections/problem_definition.tex
\section{Problem Statement}
\label{sec:problem_statement}
\begin{figure*}[th]
    \centering
    \begin{minipage}{0.42\textwidth}
    \begin{subfigure}[t]{\textwidth}
        \centering
        \includegraphics[width=\textwidth]{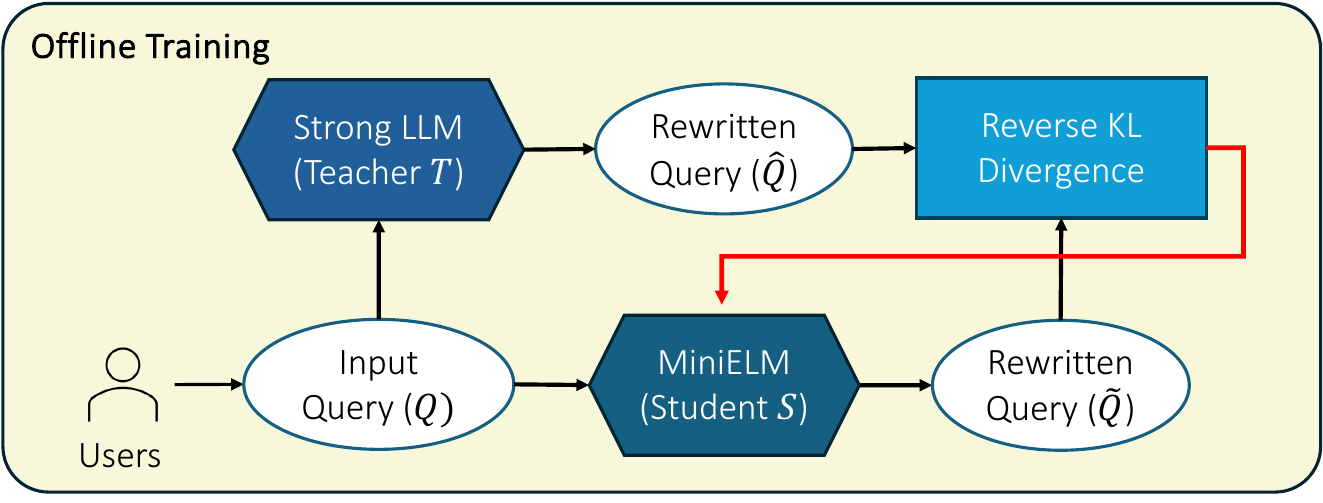}
        \caption{Offline training.}
        \label{fig:offline}
    \end{subfigure}
    
    \begin{subfigure}[t]{\textwidth}
        \centering
        \includegraphics[width=\textwidth]{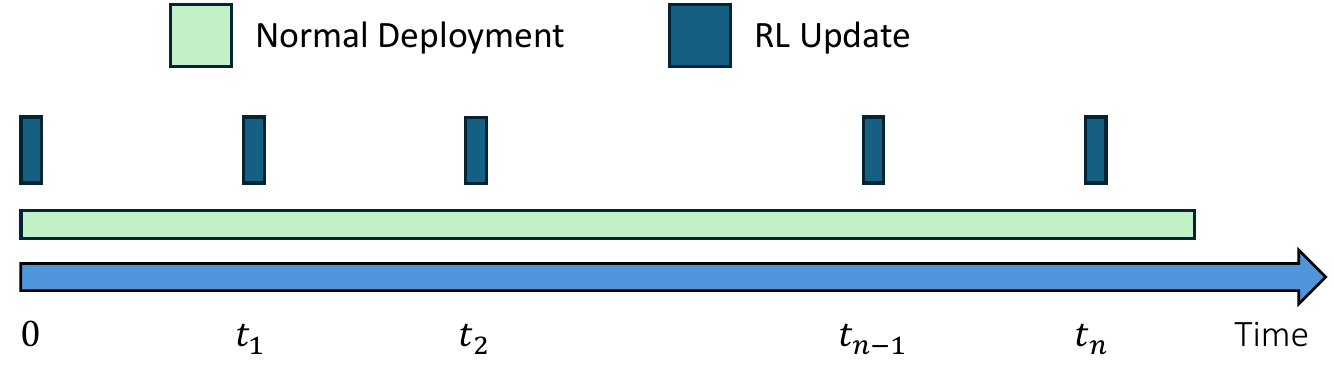}
        \caption{Simulation of online deployment.}
        \label{fig:simulation}
    \end{subfigure}
    \end{minipage}
    \hfill
    \begin{minipage}{0.55\textwidth}
    \begin{subfigure}[t]{\textwidth}
        \centering
        \includegraphics[width=\textwidth]{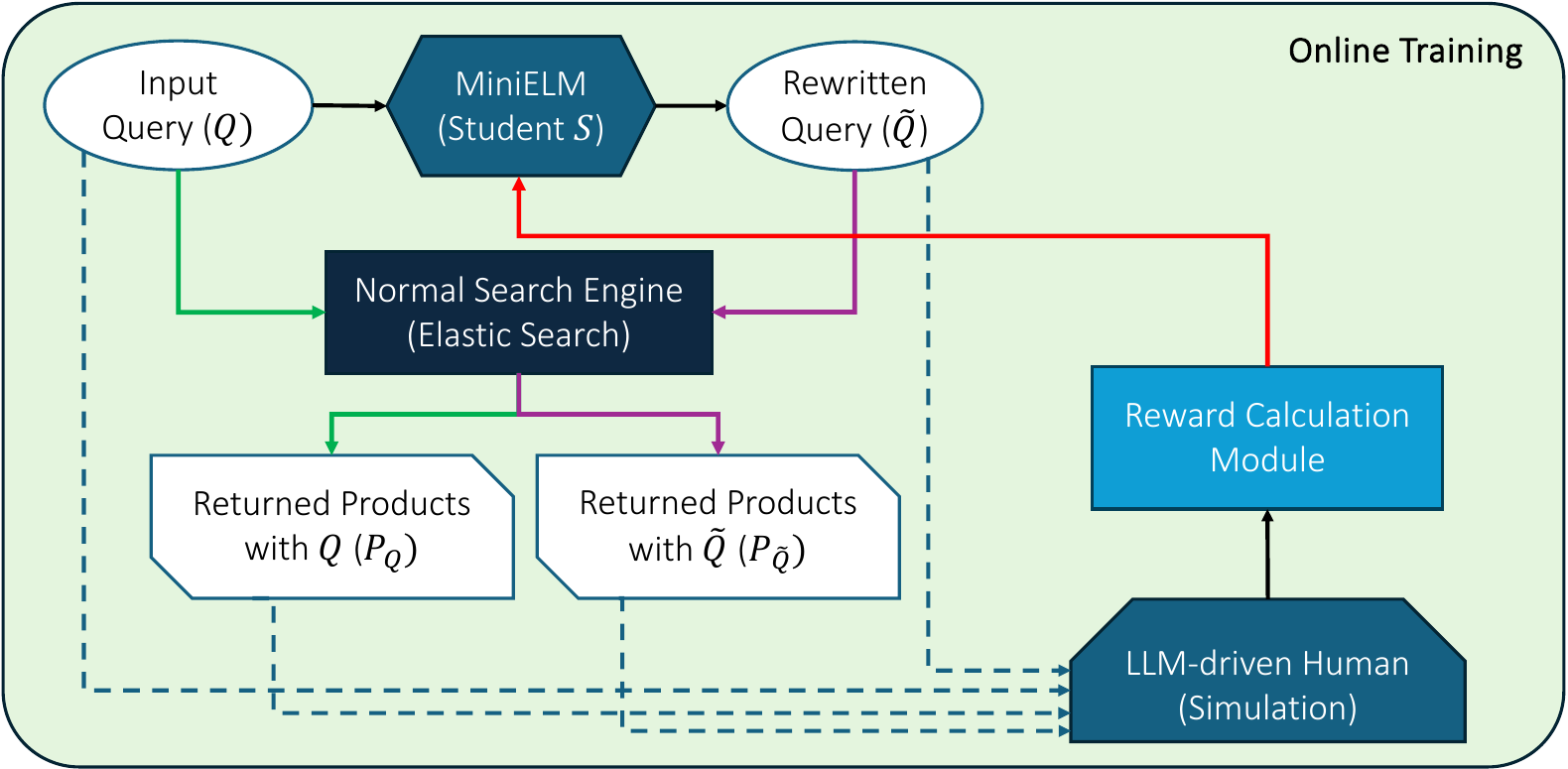} 
        \caption{Online training.}
        \label{fig:online}
    \end{subfigure}
    \end{minipage}
    \caption{High-level diagram of MiniELM's training pipelines: Offline training combines supervised fine-tuning (SFT) and knowledge distillation (KD) for a robust QR foundation, while online training leverages RL updates from custom reward signals and simulated human feedback to adapt to e-commerce dynamics.}
    \label{fig:overall}
\end{figure*}
Let $\mathcal{D} = \left\{ Q_i \right\}_{i=1}^N$ represent the dataset of real user queries collected from the historical data of e-commerce systems, where $Q_i=\left\{ t_i^1, t_i^2, \dots, t_i^{m_i} \right\}$. Here, $t_i^j$ denotes the $j^{th}$ token in the $i^{th}$ user query. The objective of the query rewriting (QR) task is to produce a corresponding set of rewritten queries, $\mathcal{Y} = \left\{ \tilde{Q}_i \right\}_{i=1}^N$, where $\tilde{Q}_i$ is the rewritten version of $Q_i$. For simplicity, we omit the index $i$ whenever the context is clear.

Since there is no definitive ground truth for an ideal rewritten query, nor should there be—this would restrict the flexibility of potential rewrites—we instead define a set of novel metrics to evaluate the quality of a rewritten query $\tilde{Q}$ relative to the original query $Q$. These metrics are computed by comparing the lists of products retrieved by the e-commerce search engine for the original query $Q$ and the rewritten query $\tilde{Q}$, denoted as $P_{Q}$ and $P_{\tilde{Q}}$, respectively. The key metrics are as follows:
\begin{itemize}
    \item \textbf{Relevant score} $r(Q, P_{\tilde{Q}})$: Measures how well the results retrieved for the rewritten query align with the intent of the original query $Q$. 
    \item \textbf{Diversity score} $d(P_Q, P_{\tilde{Q}})$: Quantifies the diversity in the product list returned for the rewritten query compared to the original.
    \item \textbf{Click/Add2cart/Purchase rate score} $c(P_{\tilde{Q}})/a2c(P_{\tilde{Q}})/p(P_{\tilde{Q}})$: Estimate the likelihood of user engagement with the product list returned for the rewritten query. These metrics simulate user behavior through Reinforcement Learning with Artificial Implicit Feedback (RLAIF) \cite{rlaif} in the online training pipeline.
\end{itemize}
Details on the calculation of these metrics, which serve both as reward signals and evaluation criteria, are provided in Section \ref{sec:method_rl}.

%% file: sections/method.tex
\section{Method}
\label{sec:method}
\input{tables/longtail}

Our approach for QR employs a dual-phased pipeline that integrates offline and online training methodologies (Figure \ref{fig:overall}). This pipeline leverages the natural language understanding of large language models (LLMs) while addressing efficiency and adaptability challenges through knowledge distillation and simulated user feedback. In the offline phase, we create MiniELM, a compact yet powerful model optimized for query rewriting, using supervised fine-tuning (SFT) on a custom Q2Q dataset and knowledge distillation (KD) to retain semantic fidelity while reducing computational overhead. This ensures MiniELM inherits the capabilities of a larger teacher model while aligning with domain-specific objectives in query rewriting. The online phase then dynamically adapts MiniELM to prioritize relevance and diversity while evolving to reflect simulated user preferences and updates in the product catalog. Together, these phases form a cohesive framework: the offline phase establishes a robust and efficient foundation, and the online phase continuously refines and personalizes the model for real-world deployment.

\subsection{Offline Training Phase}
The offline phase serves as a warm-start mechanism for the query rewriting (QR) model, ensuring that it is both \textbf{highly effective in rewriting queries} and \textbf{computationally efficient with minimal overhead}.

A key challenge in applying vanilla LLMs to e-commerce QR is their tendency to generate long-tail rewrites (as shown in Table \ref{tab:longtail}), which are often suboptimal and difficult to process in downstream search pipeline stages \cite{long_tail, long_tail1}. To mitigate this issue, we first apply supervised fine-tuning (SFT) using a curated Q2Q dataset derived from the Amazon ESCI dataset \cite{esci}. This step adapts the model to the QR task, aligning its outputs with domain-specific requirements and improving rewrite quality.
Our approach trains two model variants: a Teacher model ($T$), a large-scale LLM with strong language understanding, and a Student model ($S$), a smaller, more efficient version optimized for reduced computational overhead. 

Subsequently, a KD strategy is applied to transfer the Teacher model’s knowledge to the Student model. This two-step process - fine-tuning and distillation - ensures that the Student model inherits the Teacher’s strong performance while maintaining efficiency. Fine-tuning first allows the Teacher to learn optimal QR patterns, which are then distilled into the smaller model, preventing excessive performance degradation during compression.
The outcome of this offline training phase is MiniELM, a fine-tuned and distilled Student model that forms the foundation for the subsequent online phase.

\begin{figure}[th]
    \centering
    \includegraphics[width=0.8\columnwidth]{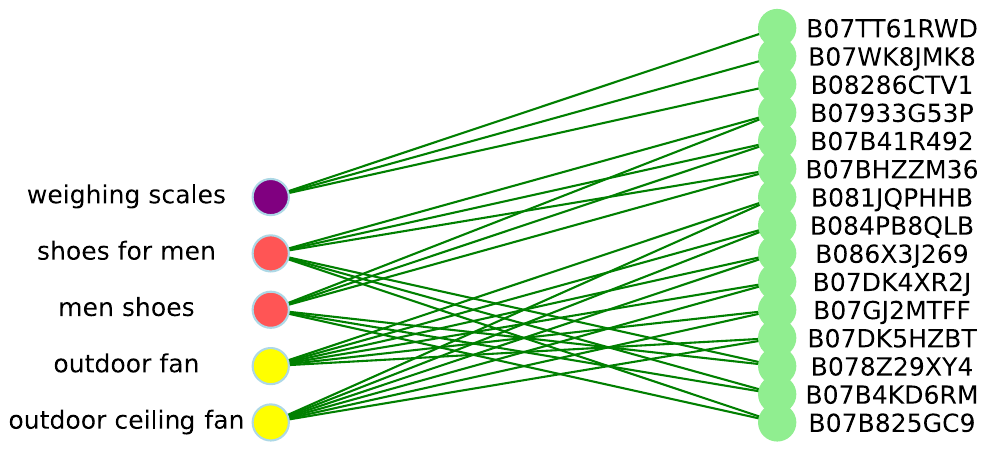}
    \caption{Illustration for Query-Product bi-partite graph.}
    \label{fig:bi_graph}
\end{figure}

\textbf{SFT with Custom Q2Q dataset.} We construct a custom query-to-query (Q2Q) dataset using existing queries from the Amazon ESCI dataset \cite{esci}. The ESCI dataset represents data as triplets ($Q, P, R$), where $Q$ is a user query, $P$ is a product in the Amazon catalog, and $R$ is the relevance score between them. Leveraging this structure, we create a bipartite graph that maps the relevance relationships between the query set and the product set (illustrated in Fig. \ref{fig:bi_graph}). From this graph, we identify query pairs that are mapped as relevant to at least $k$ similar products (e.g. labeled as ``E'' or ``S'' in the ESCI dataset). These query pairs are treated as candidate equivalents. To ensure semantic accuracy, the final set of candidate query pairs is filtered using a strong LLM (Llama 3.3, 70B version in our case), which verifies the semantic equivalence of the queries. Final selections (e.g. ``men shoes'' and ``shoes for men'' - Figure \ref{fig:bi_graph}) are then included in the custom Q2Q dataset. 
This building procedure is beneficial as it completely remove human manual annotations out of the loop, unlike existing works \cite{agrawal2023rl, long_tail}.
Building on this curated Q2Q dataset, we fine-tune both the teacher model $T$ and the student model $S$ on these query-to-query pairs. This targeted fine-tuning process ensures that both models are aligned with the task of generating accurate query rewritings within the e-commerce context. By focusing on equivalence in query rewriting, this method significantly mitigates the long-tail queries generated by vanilla LLMs. 

\textbf{KD from $T$ to $S$.} After SFT on both $T$ and $S$, an additional step of Knowledge distillation is further employed to transfer the language capabilities of $T$ to $S$. In this process, we employ the techique introduced in \cite{minillm}, with the center idea circulate around reverse Kullback-Leibler divergence (KLD) during distillation:
\begin{equation}
    \mathcal{D}_{KL}(P_S || P_T) = \sum_x P_S(x) \log \frac{P_S(x)}{P_T(x)}
\end{equation}
This loss minimizes the student model's tendency to overestimate low-probability regions of the teacher's distribution, enabling it to focus on high-relevance predictions (major modes) of $T$. 
This benefit brought about with reverse KLD is particularly favorable for generation task of $T$ or $S$ that involve a great scale dictionary, unlike normal classification tasks.

After the process, we attain the fine-tuned and distilled version of Student $S$ - MiniELM, that addresses the computational inefficiencies associated with deploying large-scale LLMs in real-time search systems, while maintaining great language ability and sense for E-Commerce QR task. 

\subsection{Online Training Phase}
\label{sec:method_rl}

The online training phase extends the offline foundation by enabling MiniELM to adapt dynamically to the e-commerce environment through real-time learning during deployment process (Figure \ref{fig:simulation}). This phase employs reinforcement learning (RL) to fine-tune the model using gradient policy optimization \cite{ppo, dpo}, ensuring that MiniELM remains responsive to updates in product catalogs and user behavior.

\textbf{Online reward signal.} 
 To effectively guide this real-time learning, the online training phase relies on carefully designed reward signals (defined in Section \ref{sec:problem_statement}), which capture the multifaceted objectives of query rewriting. The \textbf{relevance} score ensures alignment between the original and rewritten queries, maintaining consistency with users' original intents. \textbf{Diversity} measures the extent to which the rewritten query expands product coverage by retrieving distinct items compared to the original query. While both metrics can be calculated offline and provide a baseline reward signal, they fail to capture user interest in the retrieved products - a critical indicator of query quality. To address this, an \textbf{online feedback score} is derived from simulated user interactions using a judge model - named $M_2$. This score, combined with relevance and diversity, ensures the model balances query expansion with relevance to user preferences and broader exploratory needs. All of these metrics are quantified as follow.

\begin{itemize}
    \item \textbf{Relevant score} $r(Q, P_{\tilde{Q}})$: We begin by fine-tuning a \texttt{bert-base-uncased} model $M_1$ \cite{bert} on $(Q,P,R)$ pairs from the ESCI dataset to evaluate the relevance between arbitrary query-product pairs. The \textbf{relevance score} is then computed as: $r(Q, P_{\tilde{Q}})= \frac{1}{|P_{\tilde{Q}}|}\sum_i M_1 \left(Q, P_i\right) \forall P_i \in P_{\tilde{Q}}$.
    \item \textbf{Diversity score} $d(P_Q, P_{\tilde{Q}})$: This metric measures the proportion of distinct products retrieved by the rewritten query compared to the original list. It is defined as: $d(P_Q, P_{\tilde{Q}}) = \frac{|P_{\tilde{Q}}| - |P_Q \cap  P_{\tilde{Q}}|}{|P_Q|}$.
    \item \textbf{Click/Add2cart/Purchase rate score} $c(P_{\tilde{Q}})/a2c(P_{\tilde{Q}})/p(P_{\tilde{Q}})$: An LLM judge model $M_2$ is carefully prompted to assess the quality of a rewritten query based on its associated product list $P_{\tilde{Q}}$ (detailed prompts are covered in Appendix \ref{sec:app_simulation}). The model takes as input the simulated user’s bio information, drawn from a pre-synthesized profile pool (details on the pool generation process are provided in Appendix \ref{sec:app_simulation}), along with the original query $Q$ and the product list $P_{\tilde{Q}}$.  It then simulates up to $k$ interactions that the user might perform with the products. User interactions are categorized into three levels of increasing interest: \textbf{clicking} ($c(P_{\tilde{Q}})$), \textbf{adding to cart} ($a2c(P_{\tilde{Q}})$) and \textbf{purchasing} ($p(P_{\tilde{Q}})$). For each product list $P_{\tilde{Q}}$, $M_2$ is prompted to separately predict the number of interactions for each category. For instance, $c(P_{\tilde{Q}})=\frac{M2(bio,Q,P_{\tilde{Q}})}{|P_{\tilde{Q}}|}$ estimates the number of products clicked, normalized by total number of products in the list. Ideally, the interaction count should reflect the quality of $P_{\tilde{Q}}$, where higher-quality rewrites yield more positive user interactions.
\end{itemize}

\textbf{Online DPO.} We chose online \textbf{Direct Policy Optimization} (DPO) \cite{dpo} as our reinforcement learning (RL) algorithm to further align our student model, as it offers significant advantages aligned with our online deployment goals. Unlike traditional RL methods, DPO does not require a pre-collected or annotated dataset. Instead, feedback from the judge model $M_2$, along with relevance and diversity metrics, serves as the reward signal, replacing the need for manual annotations (Figure \ref{fig:online}).

At each training step, a query is sampled from the query dataset $\mathcal{D}$ (here ESCI dataset) and a rewriting pair is generated based on the current policy. The judge model $M_2$ evaluates the pair by simulating user feedback and other reward signals, selecting the response with better generation quality as the \textbf{preferred output} \begin{math}\tilde{Q}^+\end{math} and the other as the \textbf{rejected output} \begin{math}\tilde{Q}^-\end{math}. The policy is then updated using the DPO loss function:
\begin{equation}
    \small
    \mathcal{L}_{DPO}(\theta) = - \frac{1}{B} \sum_i \mathcal \log \sigma\left(\beta \log \frac{\pi_\theta\left(\tilde{Q}^{+}_i \mid Q_i\right)}{\pi_{\boldsymbol{\theta}}\left(\tilde{Q}^{-}_i \mid Q_i\right)}\right)
\end{equation}
\normalsize
Here, $B$ is the mini-batch size, $\sigma$ denotes \texttt{sigmoid} function, and $\pi_{\theta}$ is the MiniELM model with trainable parameters $\theta$. The loss intuitively minimizes the negative log-likelihood of correctly predicting the preference order.

Unlike RLHF \cite{rlhf}, DPO avoids the iterative training of a separate reward model, eliminating the need for labor-intensive data collection and annotation. By directly leveraging preference pairs and optimizing a simpler loss, DPO is more lightweight and efficient, making it ideal for real-world e-commerce deployment to align our MiniELM. 


%% file: tables/longtail.tex
\begin{table*}[th]
\centering
\small
\caption{Rewrittings generated by different LLMs given user query: ``i love you through and through board book''.}
\label{tab:longtail}
\begin{tabularx}{\textwidth}{l X}
\toprule
Model        & Rewritten Query                                                                           \\ \midrule
Llama 3 8B & love board books for toddlers and young children that express deep affection and devotion \\
GPT2-large   & board books about unconditional love and family bonds                                     \\ \bottomrule
\end{tabularx}
\end{table*}

%% file: sections/experiment.tex
\section{Experiments and Results}
\label{sec:experiment}

The primary goal of our experiments is to evaluate our proposed approach using the ESCI dataset. We begin by measuring performance across three offline metrics, followed by five online signals. The experiments demonstrate how knowledge distillation (KD) enhances query rewriting capabilities in the offline phase, while reinforcement learning (RL) improves performance across the five online signal scores. Finally, we qualitatively analyze specific query rewriting tasks to highlight how the online phase further refines and improves the model.

\subsection{Experiment Setting}
\subsubsection{Dataset} 
We use two different datasets for offline and online training, both based on Amazon ESCI (\texttt{us} locale) dataset \cite{esci}.

\textbf{Offline phase dataset.} We build our custom Q2Q dataset from the training split of the Amazon ESCI dataset. Out of $74,888$ unique queries, $23,543$ query pairs are identified as equivalent after a two-step filtering process. Since the relation is non-directional, both $(Q, \tilde{Q})$ and $(\tilde{Q}, Q)$ are included. We allocate $20\%$ of the dataset for evaluation, with the rest used for training and validation. 

\textbf{Online phase dataset.} For the simulation of MiniELM's online deployment, we perform Reinforcement Learning update with the train split of ESCI dataset, while occasionally assessing the whole pipeline performance after fix number of iterations with test split of the same dataset.

\subsubsection{Metrics} 
\textbf{Offline metrics.} Since during offline training phase, we have access to rewritten queries - served as the models' ground truth, we employ existing widely-used metrics to assess models' performance: (1) \textbf{ExactMatch} checks if the response is exactly the same as the reference text; (2) \textbf{RoughL} measures the overlap between the generated response with ground truth via their longest common subsequences; (3) \textbf{XEntropy} reports the Cross Entropy loss for generating the response.

\textbf{Online metrics.} As mentioned in Section \ref{sec:problem_statement}, we have no access to ideal rewritten queries during online deployment of MiniELM. Hence, we use the set of our custom metrics for evaluation, measuring quality of rewritten results base on desired characteristics (e.g. Relevance, diversity, positive simulated human feedback).

\subsubsection{Implementation Details} 
For both offline and online training, we adopt two LLM families for training and evaluation, suggesting that MiniELM enhance the QR task performance regardless of choice for vanilla models. Two LLM families selected are widely use GPT2 models \cite{gpt2} and state-of-the-art open-source Llama 3 models \cite{llama3}. Thoughout our experiments, we chose Llama-3.1-8B-Instruct as our judge model. For simulating ordinary E-commerce search engine, Elasticsearch with default configuration is adopted.

\textbf{Offline phase.} We select different Teacher-Student pairs for two selected model families. For GPT2, GPT2-large is selected as $T$, while base version is adopted as $S$. In parallel, Llama 3.1 8B variance is selected as $T$ and $S$ is 1B variance of Llama 3.2 model. We keep the training hyper-parameters of SFT and KD process the same as \cite{minillm} for our custom Q2Q dataset. 

\textbf{Online phase.} 
We perform simulation of actual deployment and RL update with DPO mechanism \cite{dpo} for $1000$ iterations, peforming evaluation check after $50$ updates. We adopt batch size of $16$, simulating one mini-batch DPO update for every $16$ received user queries. 

\subsection{Main Results}

\subsubsection{Evolution of MiniELM via training steps}
\input{tables/offline}

\textbf{Offline Phase Result}
Table \ref{tab:offline} presents the results of the offline training phase across different backbone LLM models, where V denotes the Vanilla (untrained) model and P represents the fine-tuned model. Two key insights emerge from these results.
First, the supervised fine-tuning (SFT) process significantly enhances the performance of both the Teacher ($T$) and Student ($S$) models on the query rewriting (QR) task. A notable limitation of vanilla LLMs is their tendency to generate long-tail queries with excessive length, which complicates downstream processes in the e-commerce search pipeline \cite{long_tail1, long_tail2, long_tail} (e.g., aligning and matching with product catalogs in e-commerce databases). The SFT process effectively mitigates this issue, enabling the fine-tuned models to produce reformulations that are more concise and better aligned with the ground truth.
Second, knowledge distillation (KD) training consistently improves the performance of the Student model ($S$), narrowing its gap with the Teacher model ($T$). This outcome reinforces the rationale behind the offline training strategy, achieving the dual goals of equipping MiniELM with familiarity in the QR task while ensuring it remains efficient and adept at natural language understanding.

\input{tables/online_t}
\textbf{Online Phase Result}
We evaluate the performance of MiniELM before and after the online simulation process with both choices of backbone LLMs to assess the impact of reinforcement learning (RL) training. The results are presented in Table \ref{tab:online}.
The data reveals a clear improvement across all recorded metrics, highlighting the positive evolution of rewritten queries over the deployment period as a result of effective RL updates. Specifically, RL training not only improves the relevance and diversity of the product lists $P_{\tilde{Q}}$ retrieved using the reformulated queries $\tilde{Q}$ but also increases the positive feedback from simulated human evaluators (represented by LLMs) within the e-commerce context. This improvement is crucial in addressing the limitations observed in static models, where performance may stagnate or degrade over time without continuous updates.

\subsubsection{Comparison with existing baselines} 
\input{tables/baseline}
\input{tables/qualitative}
\textbf{Baselines.} To demonstrate the effectiveness of MiniELM in the E-commerce query rewriting task, we compare it against the following methods:
\begin{itemize}
    \itemsep0em
    
    



    \item[(i)] \textbf{Supervised} \cite{t5}: T5 model is supervisedly trained with standard beam search for inference, serving as the foundational baseline for evaluating other methods.

    \item[(ii)] \textbf{RLQR} \cite{agrawal2023rl}: Combines generative models with reinforcement learning (RL) to improve product coverage by returning more distinct relevant products. Primarily designed for offline query rewriting.
    
    \item[(iii)] \textbf{CLOVER} \cite{clover}: A diversity-focused RL algorithm that generates high-quality, diverse reformulations, optimizing for human-assessed quality.
    
    \item[(iv)] \textbf{DRQR} \cite{drqr}: An RL method using a reward function combining F1 score and Query Performance Predictor (QPP).
    
    \item[(v)] \textbf{Task-Oriented QR} \cite{taskqr}: Employs RL to maximize relevant products retrieved, reformulating queries based on initial search results.
\end{itemize}

\textbf{Setting.} We adopt a pipeline configuration similar to \cite{agrawal2023rl} using the ESCI dataset (referred to as Aicrowd in \cite{agrawal2023rl}). For the model setup, the LLM used as our MiniELM is the base variant of the T5 model \cite{t5}, while the teacher model $T$ in the offline phase is its corresponding large variant. This alternated choice of backbone LLM is similar to \cite{agrawal2023rl} configuration, ensuring minimum bias and fairness in comparison. 
The primary metric for performance evaluation is \emph{Product Coverage} ($cov(\tilde{Q})$), as defined in \cite{agrawal2023rl}. Product Coverage is determined by counting the number of distinct relevant products returned by all reformulated queries. Following \cite{agrawal2023rl}, we set the number of reformulated queries per original query to 10. Our evaluation focuses exclusively on the EN data points within the test split of the ESCI dataset.
By replicating the experimental setup and metrics, we directly leverage the results reported in \cite{agrawal2023rl}, ensuring fairness and consistency. This approach also eliminates the need to reimplement baseline methods due to the unavailability of their private source code.

\textbf{Result.} Table \ref{tab:baseline} presents the results of all evaluated methods. Notably, our MiniELM outperforms all investigated baselines, including RLQR \cite{agrawal2023rl}, which is the second-best approach, despite not being explicitly trained to maximize Product Coverage. This superior performance can be attributed to the implicit learning of Product Coverage through our Relevance and Diversity reward signals. These signals emphasize retrieving distinct yet relevant products that complement those retrieved for the original queries, highlighting the importance of diversifying results while maintaining query relevance.

\subsection{Additional Analysis} 
\begin{figure}
    \centering
    \includegraphics[width=0.8\columnwidth]{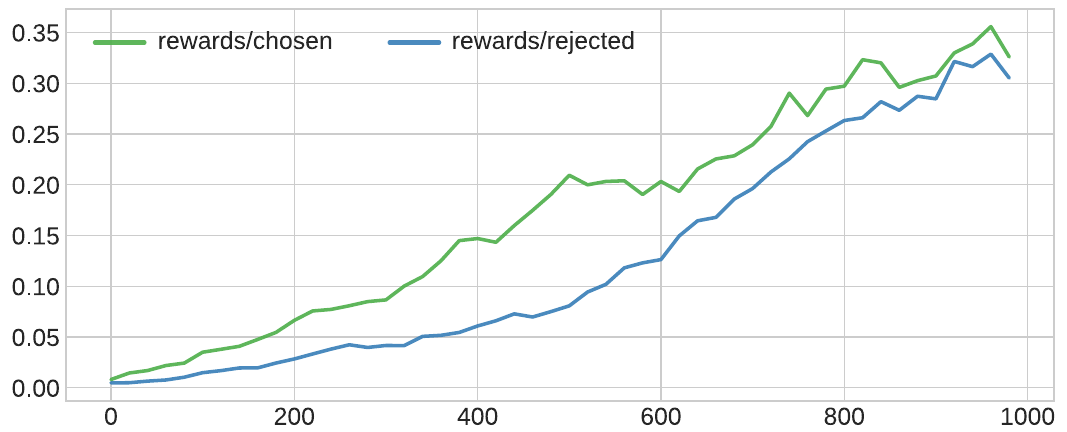}
    \caption{Rewards for both chosen and rejected rewritten queries during online RL training.}
    \label{fig:rl_training}
\end{figure}

This analysis examines MiniELM's performance evolution and query quality during the online phase.
Figure \ref{fig:rl_training} illustrates the evolution of reward signals during the online training phase using DPO for both accepted and rejected rewritten queries. The queries are generated using the MiniELM model variant with a GPT2 backbone. To highlight trends, rewards are smoothed using a $5$-window mean average.

As shown, both MiniELM's rewrites consistently improve over time, reflected in rising reward scores. This improvement highlights the effectiveness and consistency of our RL training process, demonstrating the model's ability to utilize feedback from LLMs (acting as simulated human evaluators) to refine query rewritings and enhance overall performance.

We also perform a qualitative analysis to observe how the same user queries are rewritten over time during the online training phase, with some examples summarized in Table \ref{tab:qualitative}. As training progresses, we observe that the rewritten queries increasingly include additional information. Notably, the added terms are typically generic, ensuring that the original intent of the initial queries remains preserved while enhancing their relevance and comprehensiveness.

%% file: tables/offline.tex
\begin{table}[th]

\centering
\caption{Result of different MiniELM variances on ESCI Dataset within offline training phase.}
\label{tab:offline}
\resizebox{\columnwidth}{!}{%
\begin{tabular}{@{}lllcccc@{}}
\toprule
\multicolumn{2}{l}{\textbf{Model}}                              &          & \textbf{ExactMatch}                   & \textbf{RoughL}                        & \textbf{XEntropy}                     & \textbf{Mean Length} \\ \midrule
                        &                                       & V  & 0                                     & 4.453                                  & 8.1314                                & 217.196              \\
                        & \multirow{-2}{*}{$S$}    & P & \textbf{3.125}                        & \textbf{42.256}                        & \textbf{4.632}                        & 4.265                \\ \cmidrule(l){2-7} 
                        &                                       & V  & 0.042                                 & 6.592                                  & 7.433                                 & 147.187              \\
                        & \multirow{-2}{*}{$T$}    & P & \textbf{5}                            & \textbf{44.996}                        & \textbf{4.204}                        & 9.257                \\ \cmidrule(l){2-7} 
                        &                                       & V  & 3.125                                 & 42.256                                 & {\color[HTML]{FF0000} \textbf{4.632}} & 4.265                \\
\multirow{-6}{*}{\rotatebox[origin=c]{90}{Llama 3}} & \multirow{-2}{*}{$T \rightarrow S$} & P & {\color[HTML]{FF0000} \textbf{4.5}}   & {\color[HTML]{FF0000} \textbf{43.217}} & 4.764                                 & 4.296                \\ \midrule
                        &                                       & V  & 0                                     & 0.692                                  & 9.567                                 & 213.228              \\
                        & \multirow{-2}{*}{$S$}            & P & \textbf{2.833}                        & \textbf{34.817}                        & \textbf{8.08}                         & 2.993                \\ \cmidrule(l){2-7} 
                        &                                       & V  & 0                                     & 0.831                                  & 8.454                                 & 211.98               \\
                        & \multirow{-2}{*}{$T$}           & P & \textbf{1.75}                         & \textbf{38.982}                        & \textbf{4.684}                        & 3.318                \\ \cmidrule(l){2-7} 
                        &                                       & V  & 2.833                                 & 34.817                                 & 8.08                                  & 2.993                \\
\multirow{-6}{*}{\rotatebox[origin=c]{90}{GPT2}}  & \multirow{-2}{*}{$T \rightarrow S$}         & P & {\color[HTML]{FF0000} \textbf{2.875}} & {\color[HTML]{FF0000} \textbf{35.577}} & {\color[HTML]{FF0000} \textbf{3.739}} & 3.081                \\ \bottomrule
\end{tabular}
}
\end{table}

%% file: tables/online_t.tex
\begin{table}[th]
\centering
\caption{Result of different MiniELM variances on ESCI Dataset within online training phase.}
\label{tab:online}
\resizebox{0.7\columnwidth}{!}{%
\begin{tabular}{@{}lllll@{}}
\toprule
\textbf{Metrics} & \multicolumn{2}{c}{\textbf{Llama}} & \multicolumn{2}{c}{\textbf{GPT2}} \\ \midrule
                 & $T \rightarrow S$       & RL                   & $T \rightarrow S$      & RL                  \\ \midrule
Relevant         & 0.663       & \textbf{0.707}       & 0.569       & \textbf{0.654}      \\
Diversity        & 0.769       & \textbf{0.81}        & 0.693       & \textbf{0.753}      \\
Click            & 0.513       & \textbf{0.533}       & 0.489       & \textbf{0.511}      \\
Add2cart         & 0.498       & \textbf{0.516}       & 0.466       & \textbf{0.508}      \\
Purchase         & 0.468       & \textbf{0.503}       & 0.443       & \textbf{0.502}      \\ \bottomrule
\end{tabular}
}
\end{table}

%% file: tables/baseline.tex
\begin{table}[t!]
\centering
\caption{Average relevant products returned per query on the ESCI dataset using different methods.}
\label{tab:baseline}
\resizebox{0.6\columnwidth}{!}{%
\begin{tabular}{@{}lcc@{}}
\toprule
\textbf{Method}        & \multicolumn{1}{l}{\textbf{$cov(\tilde{Q})$}} & \multicolumn{1}{l}{\textbf{Gain ($\%$)}} \\ \midrule
Supervised             & 111                                                               & 0                                     \\
RLQR                   & 145                                                               & 30.6                                  \\
CLOVER                 & 132                                                               & 18.9                                  \\
DRQR                   & 130                                                               & 17.1                                  \\
Task-Oriented QR       & 114                                                               & 2.7                                   \\ \midrule
\textbf{MiniELM (Our)} & \textbf{171}                                                               & \textbf{54.1}                                 \\ \bottomrule
\end{tabular}
}
\end{table}

%% file: tables/qualitative.tex
\begin{table*}[th]
\centering
\footnotesize
\caption{Qualitative analysis of MiniELM's rewritten queries over online training process.}
\label{tab:qualitative}
\begin{tabularx}{\textwidth}{c X X X}
\toprule
$t_0$ & red necklace                       & maternity shorts                    & boho dress 3/4 sleeve blouse             \\ \midrule
$t_1$            & red necklaces                      & women shorts                        & boho blouse dress with 3/4 sleeves       \\
$t_2$            & necklaces in red                   & mom shorts                          & boho 3/4 sleeve blouse dress             \\
$t_3$            & necklaces in red color             & maternity shorts for woman          & boho dress 3/4 sleeve blouses for women  \\
$t_4$            & red necklace for women             & comfortable maternity shorts        & boho 3/4 sleeve blouses for women        \\
$t_5$            & affordable red necklaces for women & maternity shorts for pregnant women & casual boho 3/4 sleeve blouses for women \\ \bottomrule
\end{tabularx}
\end{table*}

%% file: sections/discussion_conclusion.tex
\section{Conclusion}
\label{sec:conclusion}

This paper introduces MiniELM, a hybrid query rewriting pipeline for e-commerce that optimizes latency, cost, and adaptability. It balances performance and efficiency through offline knowledge distillation and online reinforcement learning. Experiments show improved query relevance, diversity, and user engagement. By leveraging LLM-simulated interactions, MiniELM adapts to evolving user behavior and catalogs without costly annotations, offering a scalable, cost-effective solution for dynamic e-commerce.

%% file: sections/appendix.tex
\appendix
\input{sections/prompts}

%% file: sections/prompts.tex
\section{Prompts for Human Simulation and AI Feedback Labeling}
\label{sec:app_simulation}
In this section we list the prompts we use to simulate the users' bio information and their interactions with product lists.

\textbf{Human Simulation.}
 We first defined a pool of user profiles by synthesizing their demographics (e.g., gender, age, location, income) and preferences (e.g., price sensitivity, brand affinity, style, material). By randomly sampling profiles from this pool, we simulate diverse user interactions for the same queries and product lists. The full prompt used to generate the profile pool is summarized in Table \ref{tab:human_profile}.

\begin{table*}[ht]
    \centering
    \begin{tabular}{p{0.15\textwidth} | p{0.75\textwidth}}
    \toprule
      User simulation  & Simulate the behavior of a random e-commerce user with specific demographics and preferences influencing product choices:\newline Demographics:\newline Gender: Affects preferences in apparel or cosmetics.\newline Age: Influences style, spending, and product types (e.g., 18-25, 26-35, 36-50).\newline Location: Impacts climate-related, cultural, and trending products (e.g., North America, Europe, Asia).\newline Income: Determines spending power (low, middle, high, luxury).\newline Preferences:\newline Price Sensitivity: Willingness to pay beyond budget (low to high).\newline Brand Affinity: Preference for familiar or famous brands (low to high).\newline Style: Casual, business, luxury, trendy, minimalist, or classic.\newline Material: Preference for specific or eco-friendly materials when relevant.\\ \midrule
      Task & You are now a simulated user of this ecommerce platform. \newline Choose bio and preferences for the simulated user. \\
      \bottomrule
    \end{tabular}
    \caption{Prompt used to synthesize user profile.}
    \label{tab:human_profile}
\end{table*}

\textbf{Simulating interaction.}
Given the original query $Q$ and the list of products returned by its corresponding rewritten query $\tilde{Q}$, we randomly sample a user bio to simulate their interaction with the product list $P_{\tilde{Q}}$. Table \ref{tab:interaction} shows the prompt used to simulate click behavior, with similar prompts constructed for ``add to cart'' and ``purchase'' interactions.

\begin{table*}[ht]
    \centering
    \begin{tabular}{p{0.15\linewidth} | p{0.75\linewidth}}
    \toprule
    Instruction & User Profile: \{simu\_bio\} \newline Criteria for a good list of products: 
    1. A good list of products for a query is which has accurate representation of the user intent, demographics and preferences. \newline
    2. It should have a diverse set of products matching the query. \newline
    3. It should not have products too different from the query.\newline
    4 . The main product requested (Eg. toys for kids - toys is the main product) must be given importance, not the additional clause. The additional clause must be used as a qualifier.\\ \midrule
    Task & You are now a simulated user of this ecommerce platform and want to search products using this query:\{prompt\}. \newline
    The site returns a list of product:
    \{list\_prompt\}. \newline
    Given the bio and preferences for the simulated user and based on the query, then answer this final question:
    How many items from this list will you click? Respond with a single number only, DO NOT provide other information.\\ 
    \bottomrule
    \end{tabular}
    \caption{Prompt used to synthesize click interaction.}
    \label{tab:interaction}
\end{table*}